# All-optical Compton photon source generation via modulating pulse profile by convex focusing plasma mirror

Q. Yu[1*]

*1 School of Mechanical Engineering and Rail Transit, Changzhou University, Changzhou 213164, China;*

Abstract

A relatively efficient scheme to generate energetic $\gamma$ - ray is all-optical Compton scattering based on wakefield acceleration and plasma mirror where planar or concave solid target is usually employed to reflect driving pulse. For flat plasma mirror case, the colliding pulse radius is far larger than the radius of accelerated electron beam, which leads to smaller laser utilization efficiency. The concave plasma mirror can efficiently focus reflected laser to match electron beam radius. But the produced photon beam collimation is worse due to intensive transverse field and larger transverse wave vector of colliding laser. Here, we propose a new mechanism to achieve Compton scattering via employing convex plasma mirror. The convex plasma mirror improves reflected laser longitudinal length in addition to focusing effect, which ensures longer colliding time and larger laser utilization efficiency. Importantly, it leads transverse wave vector of colliding laser to be smaller, which guarantees better photon bunch collimation.

**I. Introduction**

Plasma-based wakefield accelerator (LWFA) [1] is great promising alternatives towards a next-generation ultra-compact source of relativistic electron beams. By exploiting high power lasers and advanced LWFA technology, high-quality electron beam [2-10] have been achieved, which can be conceived to be compact particle accelerators to be used to drive ultra-compact light source, inspiring a broad range of electromagnetic radiation spectrum from the THz to x ray or $\gamma$ ray region. Based on various laser and plasma parameters, different regimes contribute to generate high-quality multi-MeV photon beams, such as the betatron radiation produced in the laser wake-field with under-dense plasma [11], the synchrotron radiation in laser-solid interaction [12], bremsstrahlung emission by laser-accelerated fast electrons interacting with the high-Z atoms [13], Compton scattering [14, 15], and in-flight positron annihilation [16]. Recently, the all-optical controllable Compton scattering have been widely concerned [17, 18], due to the routine achievement of laser-driven GeV electron beams in the laboratory allowing for the possibility of all-optical Compton sources.

Scattering an electromagnetic wave off a counter-propagation relativistic electron is named as Thomson scattering if quantum effect can be ignored or referred to as inverse Compton scattering when quantum effect comes into play. Based on the unique characters of laser plasma accelerated electrons, ultra-brightness $x$ or $\gamma$ ray with short pulse duration has been obtained by all-optical Thomson or Compton scheme [19-24]. Currently, there are two methods to produce all optical Thomson/Compton scattering, one of which utilizes two counter-propagating laser pulses where a pulse evokes wakefield to generate energetic electron beams and the other collides the accelerated electrons. The difficulty of this way is the synchronization of time and space between the colliding pulse and emitting electrons. The other manner for achieving Thomson or Compton scattering is to place a planar plasma mirror (PPM) at the end of the emitting electrons. The PPM is used to reflect the driving laser. The latter way is easier to be implemented in experiment as it just involves a pulse and a planar plasma mirror where the reflected laser naturally overlaps with energetic emitting electrons in time and space. By this method, many research groups have successfully generated high-quality $x$ or $\gamma$ ray in experiment based on plasma wake-field acceleration regime only employing a laser pulse [19-22]. K. Ta Phuoc *et al.*, produced a broadband spectrum of c $x$-rays extending up to hundreds of keV and with a 10,000-fold increase in brightness over Compton $x$-ray sources based on conventional accelerators by the combination of a laser-plasma accelerator and a plasma mirror [19]. Changqing Zhu *et al.*, systematically investigated the tabletop Compton scattering hard x-ray source and gained stable and high yield of x-ray source in hundreds of keV range from laser electron accelerator in pure nitrogen with 15 TW laser pulses [20]. Changhai Yu *et al.*, produced tunable quasi-monochromatic photon beam and improved the ray peak energy and brilliance to be MeV and : $3 \times 10^{22}$ photons $s^{-1}$ $mm^{-2}$ $mrad^{-2}$ 0.1% BW at 1 MeV [21]. Hai-En Tsai *et al.*, further optimized the parameters for generating a tunable, quasi-monoenergetic, efficient, low-background Compton backscattering (CBS) x-ray source [22]. Moreover, Compton scattering is further numerically studied to improve the emitted photon bunch quality by some theoretical physicists [23, 24]. Xing-Long Zhu et al., numerically produced bright MeV isolated attosecond (≤260 as) $\gamma$-ray pulse trains with desirable angular momentum utilizing a circularly polarized Laguerre-Gaussian laser pulse illuminating compound targets [23]. Employing the all-optical setup with a structured solid target as a reflecting foil, J F Ong *et al.*, obtained double γ-ray beams using a 1 PW laser [24].

In all-optical Compton scattering based on wakefield acceleration and planar plasma mirror, the colliding laser is the reflected laser of the driving pulse. In the wakefield regime, the driving pulse intensity generally is small in order to weaken the nonlinear effect, which leads the reflected pulse not to be enough strong and thereby limits emitted photon energy and brightness according to the estimation formula [25, 26] of emitted photon energy: $E_\gamma \approx 4\gamma^2 \eta \omega_L f(a_0)$, where $4\gamma^2$ denotes the relativistic Doppler shift. In addition, the colliding pulse radius is far bigger than emitting electron beam radius, which causes reflected laser not to fully collide accelerated electron beam so that the laser utilization efficiency is greatly lowered. To overcome it, some groups proposed a focusing plasma mirror (FPM) with concave profile to replace planar target to reflect and focus driving laser. FPM greatly improves colliding laser intensity and laser utilization efficiency. However, the collimation and emittance of emitted photon bunch will be greatly decreased due to the increased transverse electric field strength and transverse wave vector of colliding laser. In addition, the colliding laser will rapidly diverge due to tight focusing effect, which stops laser utilization efficiency from persistent increase in turn. In order to overcome it, we present a preformed focusing plasma mirror (PFPM) with convex profile instead of concave profile. The convex focusing plasma mirror will prolong colliding laser longitudinal length and decrease its transverse wave vector except for focusing effect, which causes longer colliding time, better emitted photons collimation and bigger laser utilization efficiency. In this work, we investigate the Compton scattering based on the combination of wakefield acceleration mechanism and reflecting plasma mirrors with different structures including planar mirror, concave focusing mirror and convex focusing profiles.

**II. The demonstration of laser electron colliding regime and simulation setup**

When a laser pulse penetrates a gas target, it will evoke plasma bubble to accelerate background electrons and generate energetic electron beam. When the plasma bubble reaches the reflecting target, electron acceleration terminates. By this time, the energetic electrons locate in the rear of the bubble and the driving laser locates in the front of the bubble. Then, the driving laser is reflected by solid target located in the rear of the gas target. The reflected pulse naturally overlaps and collides with forward-propagating electrons, in which the energetic $\gamma$ - photons are gained by multi-photon Compton scattering regime in our cases. The wakefield electron acceleration scheme has been researched for many years and high-quality electron beams have been obtained experimentally. So, we omit the electron acceleration process in order to save computing resource and we presuppose an electron beam with the similar feature of electrons from wakefield acceleration. The sketch diagrams of Compton scattering with PPM, FPM and PFPM reflecting targets are plotted in Fig. 1. The reflecting targets are flat, concave with parabolic profile and convex with index profile for PPM, FPM and PFPM cases as shown in Fig. 1(a)-(c). The arrows indicate the laser photons directions after reflection by different targets. The energetic electrons overlap with the reflected laser to achieve nonlinear Compton scattering and emit energetic photons. We analyze the three Compton scattering cases by theory and numerical simulations. The simulations were performed based on PIC (Particle-in-Cell) code VLPL (Virtual Laser Plasma Lab) [27]. The simulation parameters are the same in the three cases unless specified. Simulation box is $101\lambda \times 70\lambda$ in $x$ and $y$ direction with resolution of $0.03\lambda \times 0.1\lambda$. The length and radius of presupposed energetic electron bunch are $7\lambda$ and $1\lambda$ with density of $0.01n_c$. There are 4 macro-particles in each cell to stand for presupposed energetic electrons beam. The electron beam propagates from left to right with 0.5GeV peak energy and 5% energy spread. There are 8 macro-particles in each cell hinting reflecting mirror and the reflecting target density is $5n_c$. For PPM case, the reflecting mirror is flat with thickness and height of $5\lambda \times 70\lambda$. For FPM case, the reflecting plasma mirror is concave with parabolic profile of $x = k_1 y^2$ where $k_1 \approx 0.06$. In PFPM case, the reflecting mirror has convex profile of $|y| = e^{k_2 x}$ where $k_2 \approx 0.06$ and the gap (marked in Fig. 2(c)) is $L_{gap} \approx 4.2\lambda$. The laser is linearly polarized with wavelength and radius of $\lambda = 1\mu m$, $R = 10\mu m$. Normalized laser intensity is $a_0 = 1$ and half pulse duration is $15\lambda$.

### III. Simulation results and discussions

Different reflecting target profiles lead to different reflecting pulse profiles. For the plane target, the reflected laser has almost the same profile with the driving laser. In the FPM case, the reflected laser intensity enhances due to the focusing effect of the concave mirror. Moreover, the laser longitudinal length is shortened as demonstrated in Fig. 2(b) according to theoretical analyses of the path of reflecting propagating. In our case, the concave mirror is parabolic profile, so the point B' overlaps with point A. For other concave mirror with non-parabolic profile the point B' cannot overlap with point A. Analyzing the reflected laser optical path, one knows convex focusing mirror increases reflected laser longitudinal length. The simulative laser profiles and electric field distributions along $y=0$ before and after reflecting for the three cases are presented in Fig. 2(d)-(f), respectively. Fig. 2(d)-(f) demonstrate that the numerical results prove the theoretical prediction. The reflected laser longitudinal length is unchanged in the PPM case and it decreases and increases in FPM and PFPM cases compared with the original pulse length. We can get the reflected pulse longitudinal length for PPM, FPM and PFPM cases by theoretical analyses. For PPM case,

$$\tau_{reflected}^{PPM} = \tau_{incident}^{PPM} \tag{1}$$

where $\tau_{reflected}^{PPM}$ and $\tau_{incident}^{PPM}$ are the pulse length of reflected and incident laser. We employ FPM target with profile $x=k_1 y^2$ and then the reflected pulse length $\tau_{reflected}^{FPM}$ can be expressed as:

$$\tau_{reflected}^{FPM} \approx 0 \tag{2}$$

For PFPM case, we adopt target with profile of $|y|=e^{k_2 x}$. For the upper part, it is $y=e^{k_2 x}$, and then the reflected pulse length $\tau_{reflected}^{PFPM}$ can be written as:

$$\tau_{reflected}^{PFPM} = \frac{lnR}{k_2} + R \cdot \tan\left(2\theta_2 - \frac{\pi}{2}\right) \tag{3}$$

where $R$ is the incident laser radius and $\theta_2 = arctan(k_2 \cdot R)$. The simulative value $\tau_{reflected}^{PFPM} \approx 30\lambda$ is similar with the theoretical value $\tau_{reflected}^{PFPM} \approx 33\lambda$, which proves the theoretical model.

Both FPM and PFPM targets focus reflected laser, the reflected laser will defocus rapidly after focusing point for FPM case. However, the focusing effect lasts a long period for PFPM case as depicted in Fig. 3 where reflected laser just finishes focusing at $55T_0$ ($T_0$ is laser period) as demonstrated in Fig. 3(a) and it still keeps better focusing structure at $80T_0$ as shown by Fig. 3(d). That is, the focusing profile of reflected laser can maintain more than $25T_0$. The colliding laser duration is about $30\lambda$. Both electrons and laser velocity are light speed, so the theoretical colliding time is about $15T_0$. Therefore, during laser electron colliding, the colliding pulse is able to keep focusing structure, which ensures laser electron colliding to be highly effective.

Compared with PPM target, FPM and PFPM targets can well focus reflected laser to decrease laser radius and increase laser intensity. We contrast the colliding laser electric field strength felt by the emitting electrons during laser electron colliding for the PPM, FPM and PFPM cases in Fig. 4(a)-(c). FPM and PFPM targets focus reflected laser to reach $a_0 \approx 2.6$ and $a_0 \approx 2.0$ from original $a_0=1.0$. The focusing effect is weaker in PFPM case than in FPM case. It is because the reflected laser has larger longitudinal length in the former case. The electrons can experience maximum colliding field for about $5T_0$ for PPM case. But in the FPM case, it is greatly less than $5\ T_0$ due to the reduction of laser longitudinal length and rapid divergence of colliding laser. However, PFPM target increases the time interval to $10\ T_0$ because of the increase of laser longitudinal length and the long-playing maintaining of focusing structure. To further contrast laser-electron interaction time, we show the time evolutions of total energy of electrons and photons in Fig. 4(d)-(f) for above three cases. In all cases, electrons transfer their energy to photon beam, which causes electrons energy to decrease and photon bunch energy to increase during colliding. From the time evolution of total electron energy, we can see the laser electrons interaction time is about $25T_0$ for PPM case. It decreases to $15\ T_0$ in FPM case because of short colliding pulse duration. However, the PFPM increases the interaction time to more than $25T_0$ because of the extended colliding pulse duration. It must lead obtained photon energy contained in the bunch and laser utilization efficiency to be different in the three cases. The generated $\gamma$- ray beam includes $6.4\times 10^{-7} J$ energy in PPM case. The FPM target increases it to $9.6\times 10^{-7} J$ and PFPM target further enhances it to $1.28\times 10^{-6} J$. Compared with planar plasma mirror case, the laser energy conversion efficiency to $\gamma$- ray beam increases 100% by our proposed preformed focusing plasma mirror scheme. To demonstrate the effect of reflecting mirror profile on the photons number and collimation, we present the time evolutions of photons number and angle for PPM, FPM and PFPM cases in Fig. 4(g)-(i). The photon number increases with time during colliding and then reaches saturation. We gain $7.7 \times 10^7$ photons in PPM case finally. The photon number increases to $8.1 \times 10^7$ by the FPM target and it furtherly increases to about $12 \times 10^7$ in PFPM case due to the extension of reflected laser duration. It is important that the obtained photons emitting angle is lowered greatly by the proposed PFPM target. The emitting angle is about 0.028 in PPM case, but it increases to about 0.06 in FPM case. However, the proposed PFPM target decreases it to 0.0185 yet.

In Figs. 5(a)-(c), we demonstrate the angle-energy distribution for planar plasma mirror (PPM), focusing plasma mirror (FPM) and preformed focusing plasma mirror (PFPM) cases. It is evident that photon bunch has more compact distribution in angle in our improved focusing plasma mirror case compared with traditional planar plasma mirror or concave focusing plasma mirror cases. To further compare the photons collimation, we plot the obtained photon beam angle spectra for the three cases in Fig. 5(g)-(i). The proposed PFPM target effectively improves the angle spread and the angle spread decreases 2 times and 7 times than in PPM and FPM cases. Meanwhile, the angle divergence reduces more than 3 times and 10 times than in PPM and PFM cases. To compare the photons energy distribution for the three cases, we also present photon energy spectra as shown in Fig. 5(d)-(f). The peak energies are 0.54GeV, 0.63GeV and 0.73GeV for the PPM, FPM and PFPM cases. The PFPM target also reduces the photon energy spread to 20.5% from 41.4% of FPM case.

Compared with planar plasma mirror (PPM)), traditional focusing plasma mirror (FPM) lowers emitted photons collimation and the preformed focusing plasma mirror (PFPM) improves emitted photons collimation. It is because the intense focusing effect caused by FPM target immensely increases colliding laser transverse field, which enhances scattering electrons transverse velocity. As a result, emitted photon transverse velocity accordingly enlarges. Even though the reflected laser is focused as well in the PFPM case, the focusing effect is weaker, which leads to the scattering electrons and emitted photons transverse velocity to be smaller. Moreover, we find the FPM target increases transverse wave vector of colliding laser and PFPM target reduces it compared with PPM target case as exhibited in Fig. (6) where we display the colliding laser transverse electric field distributions in wave-vector space for above three cases. Larger laser transverse wave-vector brought by FPM target leads emitted photons collimation to be worse. Meanwhile, the compact distribution of laser transverse wave-vector tending to smaller value led by PFPM target causes emitted photon bunch collimation to be better.

According to Eq. (3), one knows the reflected laser duration is determined by parameter $k_2$ for PFPM case. We prove it with a couple of simulations where $k_2$=0.06 and $k_2$=0.1. The theoretical pulse duration should be $\tau_{reflected}^{PFPM} \approx 33\lambda$ and $\tau_{reflected}^{PFPM} \approx 23\lambda$. The simulative results are $\tau_{reflected}^{PFPM} \approx 30\lambda$ (as shown by Fig. 7(a)) and $\tau_{reflected}^{PFPM} \approx 20\lambda$ (as demonstrated by Fig. 7(b)), all of which match well with theoretical results. The other critical parameter for improving colliding efficiency is the colliding laser radius. We perform a couple of simulations with $L_g$=4.2$\lambda$ and $L_g$=8.6$\lambda$ as depicted in Fig. 8(a) and (b) to investigate the dependence of reflected laser radius on reflecting PFPM parameters. We find the reflected laser radius $R_{reflected}$ is proportional to the reflecting target gap $L_g$ and can be expressed as:

$$R_{reflected}=k_0 L_g \quad (4)$$

In order to gain the proportionality coefficient $k_0$, we perform 5 simulation cases with different $L_g$ and get 5 reflected pulses of different radius. We present the relation between reflected laser radius and $L_g$ on Fig. 8(c). Linearly fitting the simulative results, one can get $k_0 = 0.89$ and

$$R_{reflected}=0.89L_g \tag{5}$$

**IV. The theoretical analyses on plasma density for wakefield acceleration**

The scattering electrons are from wakefield acceleration regime. We hope accelerated electrons collide with reflected laser in the focusing surface after reflecting laser just finishes focusing. We suppose the dephasing length is just reached before laser-electron colliding. In this case, the accelerated electrons locate in the center of the bubble and the distance between laser and electrons is $\frac{1}{2}\lambda_p$. The length between focusing point to the inner vertex of the PFPM should be $\frac{1}{4}\lambda_p$, that is $\frac{1}{2}\tau_{reflected}^{PFPM}$. Therefore,

$$\frac{1}{4}\lambda_p = \frac{1}{2}\tau_{reflected}^{PFPM} \tag{6}$$

One knows $\tau_{reflected}^{PFPM} = \frac{lnR}{k_2} + R \cdot \tan(2\theta_2 - \frac{\pi}{2})$ according to Eq. (3). Substituting it and $\lambda_p$ into Eq. (6), one gets the plasma density for wakefield acceleration:

$$n_e = \begin{cases} \dfrac{n_c\lambda^2}{16\pi^2\left[\dfrac{\ln R}{k_2} + R \cdot \tan\left(2\theta_2 - \dfrac{\pi}{2}\right)\right]^2} & \text{in linear regime} \\[2ex] \dfrac{a_0 n_c\lambda^2}{16\pi^4\left[\dfrac{\ln R}{k_2} + R \cdot \tan\left(2\theta_2 - \dfrac{\pi}{2}\right)\right]^2} & \text{in nonlinear regime} \end{cases} \tag{7}$$

The reflected laser by PFPM will not rapidly diverge. In this case, it is better to achieve laser electrons colliding in the head of the pulse after reflecting laser just finishes focusing. In this case,

$$\frac{1}{2}\lambda_p = \frac{3}{2}\tau_{reflected}^{PFPM} \tag{8}$$

In this way, the plasma density for wakefield acceleration in PFPM case should satisfy the following condition:

$$n_e = \begin{cases} \dfrac{n_c \lambda^2}{36\pi^2 \left[\dfrac{\ln R}{k_2} + R \cdot \tan\left(2\theta_2 - \dfrac{\pi}{2}\right)\right]^2} & \text{in linear regime} \\ \dfrac{a_0 n_c \lambda^2}{36\pi^4 \left[\dfrac{\ln R}{k_2} + R \cdot \tan\left(2\theta_2 - \dfrac{\pi}{2}\right)\right]^2} & \text{in nonlinear regime} \end{cases} \quad (9)$$

When plasma density employed for wakefield acceleration, laser intensity and PFPM parameter $k_2$ satisfy Eq. (7), accelerated electrons and reflected laser will collide at the pulse focusing point when reflecting pulse just realizes focusing. Energetic electrons will scatter reflected pulse at the head of the pulse when reflecting pulse just realizes focusing if Eq. (9) is met. In both cases, energetic electrons can well scatter the reflected laser. In our PFPM case, it is better to achieve electrons laser colliding at the head of the focusing laser to increase colliding time, because the reflected laser will not rapidly diverge.

**V. Conclusion**

In conclusion, we propose a preformed convex plasma mirror to improve Compton scattering based on wakefield acceleration scheme. Compared with flat plasma mirror, we find it can well focus reflected laser pulse to small radius to match the scattering electron beam radius, which improves scattering efficiency. Moreover, the pulse length of reflected laser is extended and it increases colliding time. It furtherly improves scattering efficiency and photon bunch brightness. Importantly, the convex plasma mirror decreases reflected laser transverse wave vector, which improves emitted photon bunch collimation. Focusing colliding laser to match scattering electron beam radius can also be realized by concave focusing plasma mirror. But the colliding laser will rapidly diverge after focusing point due to tightly focusing effect, which decreases Compton scattering efficiency. Moreover, the colliding laser transverse wave vector increases in concave plasma mirror case and it deteriorates emitted photon collimation. The Compton scattering efficiency increases to 200% and 133% and the collimation improves 50% and 30% in convex focusing plasma mirror case compared with plane plasma mirror and concave focusing plasma mirror cases.

We also research the effects brought by convex mirror parameters on reflected laser profile by theory and simulations. And we display the dependences of reflected laser length and radius on convex mirror parameter $k_2$, $L_g$ and incident laser radius. Considering wakefield acceleration, we present the dependence of plasma density of wakefield acceleration on driving laser intensity and convex mirror parameter $k_2$ by theoretical analyses as well, with which the driving laser focusing and colliding with energetic electrons can effectively proceed.

**Data availability statements**

The data that support the findings of this study are available upon reasonable request from the authors.

**Acknowledgements**

This work is supported by Natural Science Foundation of China under Contract No. 11804348.

# Figure Captions

**Figure 1**
The sketch diagrams of Compton scattering based on bubble electron acceleration regime with PPM, FPM and PFPM reflecting targets are shown in (a), (b) and (c). The reflecting targets are plane, concave with parabolic outline and convex with index profile, respectively. The arrows indicate the laser photons directions before and after of reflecting by the targets.

**Figure 2**
The analysis diagrams of laser path before and after reflecting are plotted in (a)-(c) for PPM, FPM and PFPM cases. Drawings (d)-(f) show the simulative reflected laser transverse electric field distributions and compare them with laser before reflecting.

**Figure 3**
Electric field distributions of reflected laser in convex focusing plasma mirror case at $55T_0$, $65T_0$, $75T_0$ and $80T_0$ are plotted in (a)-(d).

**Figure 4**
The time evolutions of maximum transverse electric field felt by the scattered electrons during laser electron colliding for PPM, FPM and PFPM cases are depicted in (a)-(c).

**Figure 5**
Energy and angle correlation of scattered photons for PPM, FPM and PFPM cases are drawn in (a)-(c). The scattered photon energy and angle spectra for PPM, FPM and PFPM cases are shown in (d)-(f) and (g)-(i).

**Figure 6**
The transverse electric field distributions of reflected laser on wave-vector space for PPM, FPM and PFPM cases are plotted in (a)-(c).

**Figure 7**
The transverse electric field distributions of reflected laser by convex plasma mirror with k=0.06 and k=0.1 are depicted in (a) and (c). (b) and (d) show the corresponding electric field distributions along $y = 35\lambda$.

**Figure 8**
The transverse electric field distributions of reflected laser by convex plasma mirror with plasma mirror gap $r_{gap}$=4.2 $\lambda$ and $r_{gap}$ =8.6 $\lambda$ are demonstrated in (a) and (b). The dependence of reflected laser radius $r_{laser}$ on the convex plasma mirror gap $r_{gap}$ is shown in (c).

**Figure 1**

**All-optical Compton photon source generation via modulating pulse profile by convex focusing plasma mirror**

**Q. Yu**

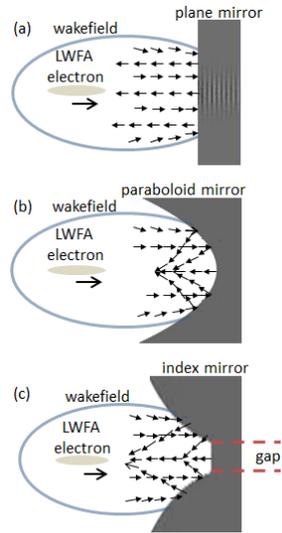

**Figure 2**

**All-optical Compton photon source generation via modulating pulse profile by convex focusing plasma mirror**

**Q. Yu**

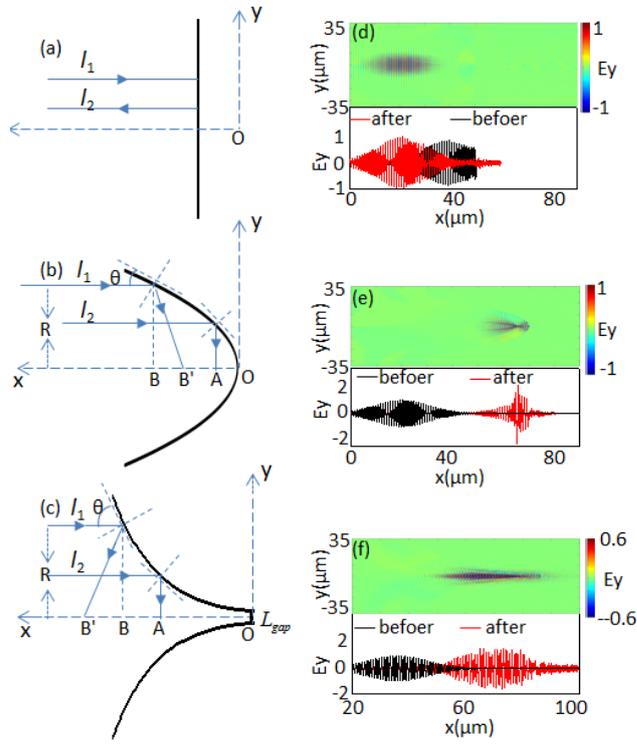

**Figure 3**

**All-optical Compton photon source generation via modulating pulse profile by convex focusing plasma mirror**

**Q. Yu**

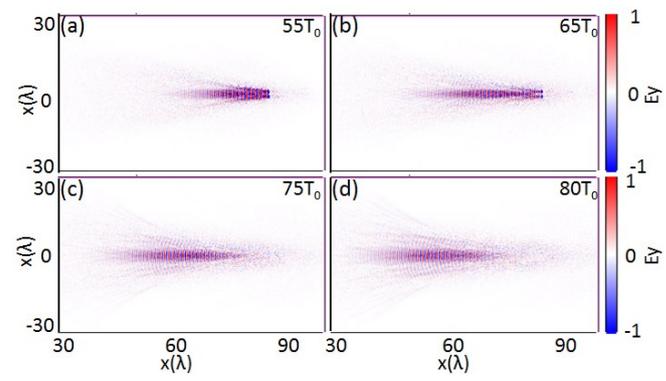

**Figure 4**

**All-optical Compton photon source generation via modulating pulse profile by convex focusing plasma mirror**

**Q. Yu**

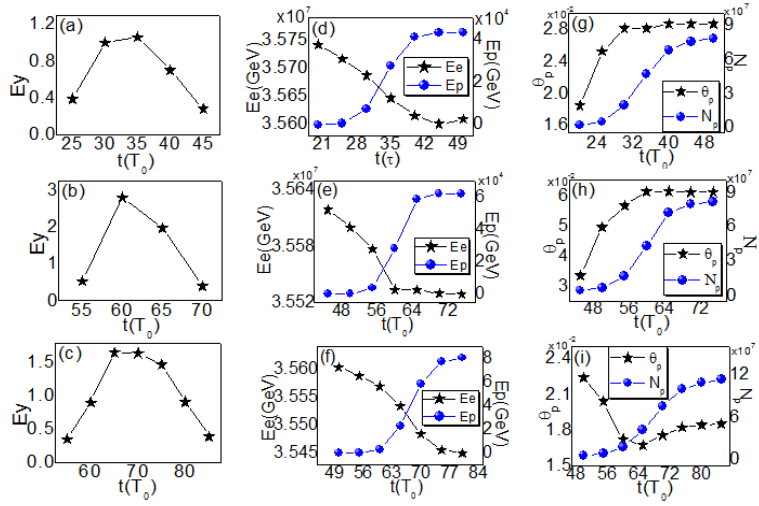

**Figure 5**

**All-optical Compton photon source generation via modulating pulse profile by convex focusing plasma mirror**

**Q. Yu**

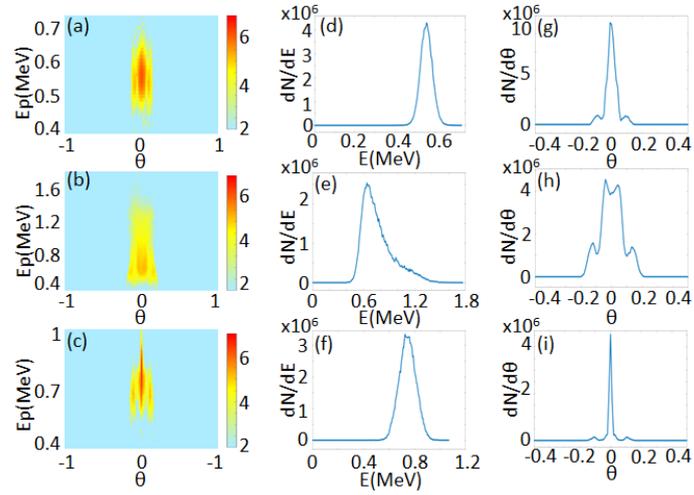

**Figure 6**

**All-optical Compton photon source generation via modulating pulse profile by convex focusing plasma mirror**

Q. Yu

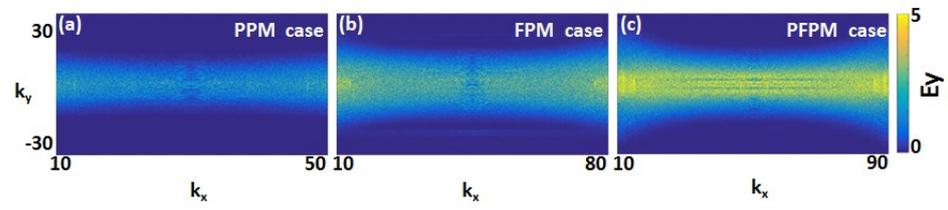

**Figure 7**

**All-optical Compton photon source generation via modulating pulse profile by convex focusing plasma mirror**

**Q. Yu**

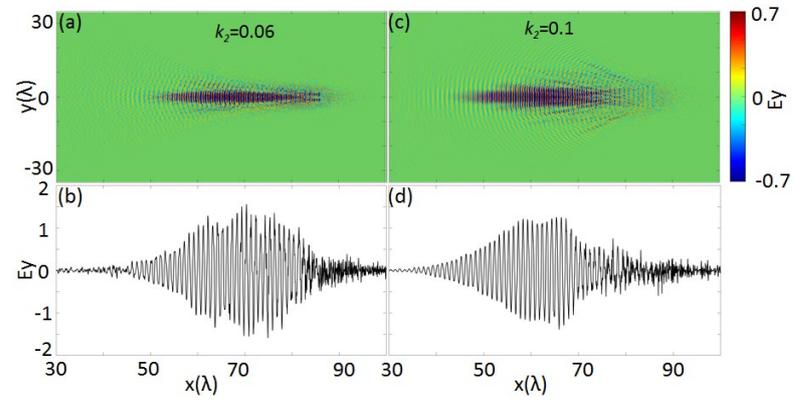

**Figure 8**

**All-optical Compton photon source generation via modulating pulse profile by convex focusing plasma mirror**

Q. Yu.

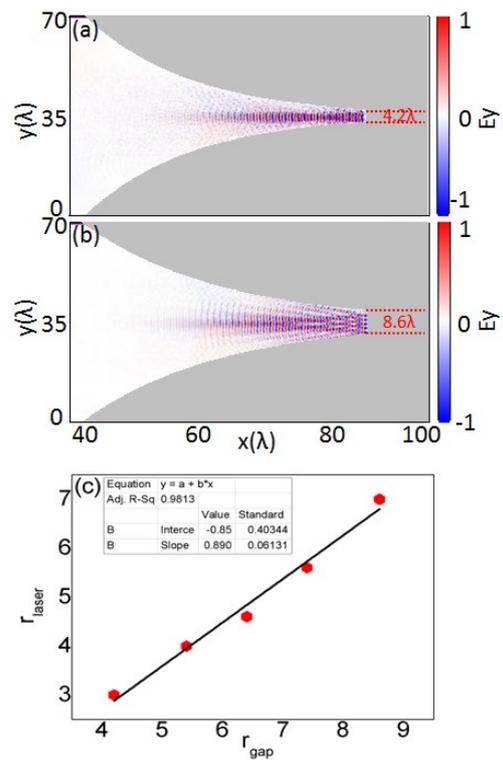